\documentclass[a4paper,11pt]{article}
\usepackage{aaskaiid}
\usepackage{units}
\usepackage{xspace}
\usepackage{graphicx}
\usepackage{subcaption}
\usepackage{orcidlink}

\newcommand{\Xmax}{$X_{\rm max}$\xspace}
\newcommand{\Xmaxmath}{X_{\rm max}}

\title{Anomalous Air Showers and What They Reveal About Hadronic Interactions and Cosmic-ray Masses}
\ShortTitle{Anomalous air showers}

\author*[d,e]{Stijn~Buitink\orcidlink{0000-0002-6177-497X}}
\author[d]{Vital~De Henau\orcidlink{0009-0003-0337-3558}}
\author[b]{Sjoerd~Bouma\orcidlink{0000-0002-6959-2302}}
\author[c]{Justin~Bray\orcidlink{0000-0002-0963-0223}}
\author[d,e]{Arthur~Corstanje\orcidlink{0000-0001-5992-6228}}
\author[f]{Edwin~Dickinson\orcidlink{0000-0003-0834-4708}}
\author[j,k]{Brian~Hare\orcidlink{0000-0001-5138-1235}}
\author[a]{Andreas~Haungs\orcidlink{0000-0002-9638-7574}}
\author[l,v]{Haoning~He\orcidlink{0000-0002-8941-9603}}
\author[e,d,m]{J\"org~H\"orandel\orcidlink{0000-0001-6604-547X}}
\author[a,d]{Tim~Huege\orcidlink{0000-0002-2783-4772}}
\author[f]{Clancy~James\orcidlink{0000-0002-6437-6176}}
\author[b]{Philipp~Laub\orcidlink{0009-0003-2617-9109}}
\author[l]{Xingyu Li\orcidlink{}}
\author[a]{Hermann-Josef~Mathes\orcidlink{}}
\author[e,m]{Katharine~Mulrey\orcidlink{0000-0001-8026-8020}}
\author[b,n]{Anna~Nelles\orcidlink{0000-0002-1720-6350}}
\author[a, x]{Subhadip~Saha\orcidlink{0000-0003-2435-8317}}
\author[w]{Felix~Schl\"uter\orcidlink{0000-0002-5545-4363}}
\author[j]{Olaf~Scholten\orcidlink{0000-0003-3649-1254}}
\author[c]{Ralph~Spencer\orcidlink{0009-0009-6015-1787}}
\author[k]{Christopher~Sterpka\orcidlink{0000-0001-8217-0836}}
\author[b]{Karen~Terveer\orcidlink{0009-0002-9594-0419}}
\author[p]{Satyendra~Thoudam\orcidlink{0000-0002-7066-3614}}
\author[q]{Gia~Trinh\orcidlink{0000-0002-5352-5092}}
\author[k]{Paulina~Turekova\orcidlink{0009-0006-1262-7507}}
\author[a]{Darko~Veberic\orcidlink{0000-0003-2683-1526}}
\author[a]{Keito~Watanabe\orcidlink{0000-0003-0599-4035}}
\author[s,t]{Chao~Zhang\orcidlink{0000-0001-9366-0056}}
\author[u]{Pengfei~Zhang\orcidlink{0000-0002-6855-5315}}
\author[l]{Yi~Zhang\orcidlink{0000-0001-6223-4724}}

\ShortName{Stijn Buitink et al.} 

\newcommand{\affilASTRON}{Netherlands Institute for Radio Astronomy (ASTRON), Dwingeloo, The Netherlands}
\newcommand{\affilCanTho}{Physics Education Department, School of Education, Can Tho University, Campus~II, 3/2 Street, Ninh Kieu District, Can Tho City, Viet Nam}
\newcommand{\affilCurtin}{International Centre for Radio Astronomy Research, Curtin University, Bentley, 6102, WA, Australia}
\newcommand{\affilDESY}{Deutsches Elektronen-Synchrotron DESY, Platanenallee~6, 15738 Zeuthen, Germany}
\newcommand{\affilErlangen}{Erlangen Centre for Astroparticle Physics, Friedrich-Alexander-Universit\"at Erlangen-N\"urnberg, 91058 Erlangen, Germany}
\newcommand{\affilGorlitz}{Deutsches Zentrum f\"ur Astrophysik, Postplatz~1, 02826 Görlitz, Germany}
\newcommand{\affilGroningen}{Kapteyn Astronomical Institute, University of Groningen, P.O.~Box 72, 9700 AB Groningen, Netherlands}
\newcommand{\affilHefei}{School of Astronomy and Space Science, University of Science and Technology of China, Hefei 230026, China}
\newcommand{\affilKanpur}{Department of Physics, Indian Institute of Technology Kanpur, Kanpur, UP-208016, India}
\newcommand{\affilKeyNanjing}{Key Laboratory of Modern Astronomy and Astrophysics, Nanjing University, Ministry of Education, Nanjing 210023, China}
\newcommand{\affilKIT}{Institut f\"ur Astroteilchenphysik, Karlsruhe Institute of Technology (KIT), P.O.~Box 3640, 76021 Karlsruhe, Germany}
\newcommand{\affilKhalifa}{Department of Physics, Khalifa University, P.O.~Box 127788, Abu Dhabi, United Arab Emirates}
\newcommand{\affilManchester}{Jodrell Bank Centre for Astrophysics, Department of Physics and Astronomy, University of Manchester, Manchester M13 9PL, UK}
\newcommand{\affilMaxPlanck}{Max-Planck Institut f\"ur Astrophysik, Karl-Schwarzschild-Str.~1, 85748 Garching, Germany}
\newcommand{\affilMunich}{Ludwig-Maximilians-Universit\"at M\"unchen (LMU), Geschwister-Scholl-Platz~1, 80539 M\"unchen, Germany}
\newcommand{\affilNanjing}{School of Astronomy and Space Science, Nanjing University, Nanjing 210023, China}
\newcommand{\affilNijmegen}{Department of Astrophysics/IMAPP, Radboud University Nijmegen, P.O.~Box 9010, 6500 GL Nijmegen, The Netherlands}
\newcommand{\affilNikhef}{Nikhef, Science Park Amsterdam, 1098 XG Amsterdam, The Netherlands}
\newcommand{\affilPurpleMt}{Key Laboratory of Dark Matter and Space Astronomy, Purple Mountain Observatory, Chinese Academy of Sciences, No.~10 Yuanhua Road, Nanjing, China}
\newcommand{\affilULB}{Universit\'e Libre de Bruxelles, Science Faculty CP230, B-1050 Brussels, Belgium}
\newcommand{\affilVUB}{Inter-University Institute For High Energies (IIHE), Vrije Universiteit Brussel (VUB), Pleinlaan 2, 1050 Brussels, Belgium}
\newcommand{\affilXidian}{School of Electronic Engineering, Xidian University, No.2 South Taibai Road, Xi'an, China}
\newcommand{\affilSKA}{SKA Observatory, Jodrell Bank, Lower Withington, Macclesfield, SK11 9FT, UK}
\newcommand{\affilIITK}{Department of Physics, Indian Institute of Technology Kanpur, Kanpur, UP-208016, India}

\affiliation[a]{\affilKIT}
\affiliation[b]{\affilErlangen}
\affiliation[c]{\affilManchester}
\affiliation[d]{\affilVUB}
\affiliation[e]{\affilNijmegen}
\affiliation[f]{\affilCurtin}
\affiliation[g]{\affilMaxPlanck}
\affiliation[h]{\affilMunich}
\affiliation[i]{\affilGorlitz}
\affiliation[j]{\affilGroningen}
\affiliation[k]{\affilASTRON}
\affiliation[l]{\affilPurpleMt}
\affiliation[m]{\affilNikhef}
\affiliation[n]{\affilDESY}
\affiliation[o]{\affilKanpur}
\affiliation[p]{\affilKhalifa}
\affiliation[q]{\affilCanTho}
\affiliation[r]{\affilSKA}
\affiliation[s]{\affilNanjing}
\affiliation[t]{\affilKeyNanjing}
\affiliation[u]{\affilXidian}
\affiliation[v]{\affilHefei}
\affiliation[w]{\affilULB}
\affiliation[x]{\affilIITK}

\emailAdd{stijn.buitink@vub.be}

\abstract{The identification of the sources and acceleration mechanisms of cosmic rays require precise measurements of their mass composition. Currently, the most reliable method is to measure the atmospheric depth at which cosmic ray air showers in our atmosphere reach their maximum (\Xmax). However, the hadronic interaction properties that govern the longitudinal development of air showers are not precisely known, which is a major source of systematic uncertainty on the mass composition. 

SKA-Low will observe cosmic rays in the 10$^{16}$ - 10$^{18}$ eV energy range with unprecedented resolution and bandwidth. This allows for a much more detailed reconstruction of the longitudinal shower evolution, which can be used to gain better understanding of the hadronic interactions, as well as the primary mass composition. 

After the first interaction of the cosmic ray with an atom in an air molecule, the secondary particles still carry a significant fraction of the total energy. When one of these particle travels very far before interacting again, it produces a sub-shower that can be recognized as a secondary bump in the longitudinal profile. Simulations have demonstrated that SKA-Low can resolve such double bump profiles by virtue of its high antenna density and broad bandwidth.

In this chapter, we demonstrate how double-bump showers and other anomalous longitudinal developments can be used to constrain hadronic interaction properties, and to determine the mass composition of cosmic rays in the Galactic-to-extragalactic transition region.}

\begin{document}
\maketitle

\section{Introduction}

Cosmic rays are the most energetic particles in the Universe and have been observed up to tremendous energies of 10$^{20}$ eV. Where they come from and how they are accelerated remain large open questions in astrophysics~\citep{Blumer:2009jrd}. The main reason why source identification is hard, is that cosmic rays are charged particles and are deflected in Galactic and extragalactic magnetic fields. When they reach the Earth, their arrival directions do not point back to the source. The two main observables that can lead to a better understanding of cosmic rays are their energy spectrum and the mass composition.  

The evolution of cosmic-ray mass composition as a function of energy contains crucial information about the relative contributions of different source populations, their acceleration mechanisms, and their chemical composition~\citep{KU12}. However, the challenge is to disentangle these components. At energies below 10$^{15}$~eV there is strong evidence that the cosmic-ray flux is dominated by particles accelerated through diffusive shock acceleration in Galactic sources like supernova remnants~\citep{Fermi-LAT:2013iui}. At the highest energies, the sources must be dominantly extragalactic to explain the observed high level of isotropy \cite{PierreAuger:2017pzq}. Extragalactic sources may include blazars or other active galaxies, or mergers of neutron star binaries. They are expected to dominate the observed spectrum at earth above $2 \times 10^{18}$~ eV, where the energy spectrum hardens (an observational feature known as the ankle).

An energy range of particular importance is the transition region, loosely defined as $10^{15} - 10^{18}$~eV. It is notoriously hard to understand because it is expected to harbor a transition from a Galactic to an extragalactic flux. To make things even more complicated, there is a strong case for the existence of a secondary
Galactic component, which can accelerate cosmic rays up to energies 10 to 100 times larger than the maximum energy that the main Galactic component can achieve. Possible scenarios are reacceleration on the Galactic termination shock, or shockwave acceleration in supernovas expanding into the strongly magnetised winds surrounding Wolf-Rayet stars~\citep{Thoudam:2016}. For any source that produces cosmic rays through diffusive shock acceleration, the maximum attainable energy is rigidity dependent. As a result, a trend from light to heavy in the mass composition spectrum can mark the maximum energy of a source class~\citep{Peters:1961mxb}. Such features have been found in the transition region~\citep{KASCADEGrande:2011kpw} and at the end of the cosmic-ray spectrum~\citep{PierreAuger:2024flk}. In addition, propagation of cosmic rays in the Galactic magnetic field will affect the mass composition measured at Earth.

An accurate measurement of the mass composition is needed to identify and disentangle the different source contributions in the cosmic-ray flux. Above $10^{15}$~eV this requires the observation of the particle cascades, or \emph{extended air showers}, that the cosmic rays produce in the atmosphere. Such observations can be performed with particle detectors, Cherenkov and fluorescence telescopes, or radio arrays~\citep{Huege_2016}. The most precise radio-based mass-composition studies have been performed by LOFAR~\citep{Buitink:2016nkf, Corstanje:2021kik} and the Pierre Auger Observatory~\cite{PierreAuger:2023lkx}.

However, all cosmic-ray mass composition studies suffer from a major systematic uncertainty. The relations between direct observables (e.g. the depth of the shower maximum in the atmosphere) and the mass of the primary particles are based on Monte Carlo codes in which all particle interactions inside the shower are simulated. The results of these simulations depend on how the hadronic interaction physics is modeled. The difference between state-of-the-art models is large enough to have a strong impact on the interpretation of the data.    

SKA-Low will observe air showers in the $10^{15} - 10^{18}$~eV energy range with unprecedented precision. The extremely high antenna density and broad frequency bandwidth provide a wealth of information that was previously inaccessible. This allows the development of novel data analysis strategies that image the longitudinal shower development in the atmosphere and provide new constraints on the hadronic physics in air showers, and a more model-independent determination of the mass composition.     

\subsection{Hadronic physics in air showers}

Extended air showers contain a large variety of fundamental particles. The core of the shower consists of hadrons. When the primary cosmic ray undergoes its first inelastic scattering with a nucleus in an air atom, its energy is divided over a large number of secondary hadrons, most of which are pions. Approximately one-third of these pions are neutral and decay into gamma-ray pairs. These gamma-rays produce electromagnetic showers: they create electron-positron pairs which in turn produce new gamma-rays through bremsstrahlung. The electrons and positrons are responsible for the radio emission of the air shower. 

The other two-thirds of the pions are charged. These pions will undergo further interactions with the atmosphere, dividing the total energy over an increasing number of pions. At each step, additional energy is transferred to the electromagnetic component by the neutral pions. As their energy decreases, it becomes increasingly likely for the charged pions to decay into muons before interacting with an air molecule. The muons mostly reach the Earth's surface without interacting.  

The shower reaches its maximum number of particles when the average energy of the electrons and positrons drops below 84 MeV~\citep{heitler1984quantum}. Below this energy, the ionization energy loss becomes larger than the bremsstrahlung energy loss, and the production of gamma-rays halts.

The evolution of the shower depends crucially on a number of characteristics of the hadronic interactions. The cross section for inelastic scattering determines the average depth in the atmosphere at which the first interaction takes place. In addition, it governs the average distance that the secondary pions travel before their next interaction. The inelasticity determines how much energy is transferred to the leading particle after each interaction. Finally, the multiplicity determines how many secondary particles are created in each interaction.

The most important observable for mass composition studies with radio, Cherenkov or fluorescence measurements is \Xmax, the atmospheric depth at which the shower reaches its maximum. Changing the cross section, elasticity, or multiplicity will have an effect on the average value of \Xmax~\citep{Ulrich:2010rg,KU12}. 

Unfortunately, these parameters cannot be derived from first principles. Moreover, it is hard to measure them directly at particle accelerators. Hadrons in air showers are highly relativistic particles that collide on a fixed target in the atmosphere. The secondary particles are produced in the forward direction. Directly measuring particle production in this phase space in a ring accelerator like the LHC, would require placing instruments inside the beam pipe or far away from the interaction point. The best data in the forward regime come from the LHCf detector~\citep{LHCf:2020hjf}, which is placed 141~m from the interaction point, where the beam pipe has curved far enough to have room for detectors. In the future, the Forward Physics Facility, placed at an even larger distance of 627~m from the interaction point, will perform neutrino measurements to further constrain hadron production in the forward region~\citep{Soldin:2024rdb}. 

To model hadronic interactions, several codes have been developed that are based on phenomenological descriptions and tuned to the available accelerator data. Three of these models are available for use in air shower simulations up to the highest cosmic-ray energies: EPOS~\citep{Pierog:2023ahq}, QGSJETII~\citep{Ostapchenko_2006}, and Sibyll~\citep{PhysRevD.50.5710}. These models are regularly updated and have different versions. The simulations in this study are produced with EPOS-LHC, QGSJETII-04, and Sibyll 2.3d. Newer versions are already available and contain some significant changes~\citep{Pierog:2023ahq}. They will be incorporated in the future.  

\section{Anomalous showers}
The development of extended air showers follows from the interaction of fundamental particles that are stochastic in nature. It is therefore expected that two primary cosmic rays of the same mass and energy will produce showers with a different longitudinal development. This is particularly true for the first interaction. Air showers can start their development over a wide range of altitudes. On average, protons and light nuclei will penetrate deeper in the atmosphere than heavy nuclei, but for individual air showers the primary mass can not be inferred directly by reconstruction the atmospheric depth of the first interaction. 

After a couple of interactions, the number of particles in the air shower quickly increases. Depending on the energy of the primary, the shower will eventually contain millions or billions of secondary particles. At this stage, the stochastic behavior of individual particles does no longer have an impact on the shower development. Instead, the longitudinal development of the shower now follows a predictable pattern. This is known as \emph{shower universality}~\citep{Lafebre:2009en}. The general shape of the longitudinal evolution can be understood through the Matthews-Heitler model~\citep{Matthews:2005sd}, but an accurate description requires a full Monte Carlo simulation of all particle interactions in the shower. Traditionally, the profile is fitted with a Gaisser-Hillas function~\citep{1977ICRC....8..353G}, which can parametrized in the following way~\citep{ANDRINGA2011360}:

\begin{equation}
\label{eq:RL}
N(X) = N_\mathrm{max}\exp\left(-\frac{X - \Xmaxmath}{RL}\right)\,\left(1 + \frac{R}{L}\left(X - \Xmaxmath\right)\right)^\frac{1}{R^2},
\end{equation}
where $N(X)$ is the number of shower particles at atmospheric depth $X$, which reaches its maximum value $N_\mathrm{max}$ at the depth \Xmax. Variations in the shape are described by $L$, the width of the longitudinal profile, and $R$, which is proportional to the skewness of the profile. The shape parameters for average shower profiles above 10$^{18}$~eV have been constrained by the Pierre Auger Observatory~\citep{PierreAuger:2018gfc}.

The simple picture of a random first interaction followed by a universal development suffices for the majority of air showers, but there are exceptions. The secondary particles that are created in the first interaction can still carry a very large fraction of the primary energy. If one of these particles has a rare interaction it may leave an imprint on the shower development. The most spectacular are \emph{double bump showers}. 

\subsection{Double bump showers}
A double bump shower occurs when a very energetic secondary particle travels for a very long time before interacting. When the particle finally interacts, it initiates a sub-shower that is displaced from the main shower. If the distance covered is large enough, this results in a secondary peak in the longitudinal profile~\citep{Baus:2011kc}.

\begin{figure}
    \centering
	\includegraphics[width=0.49\columnwidth]{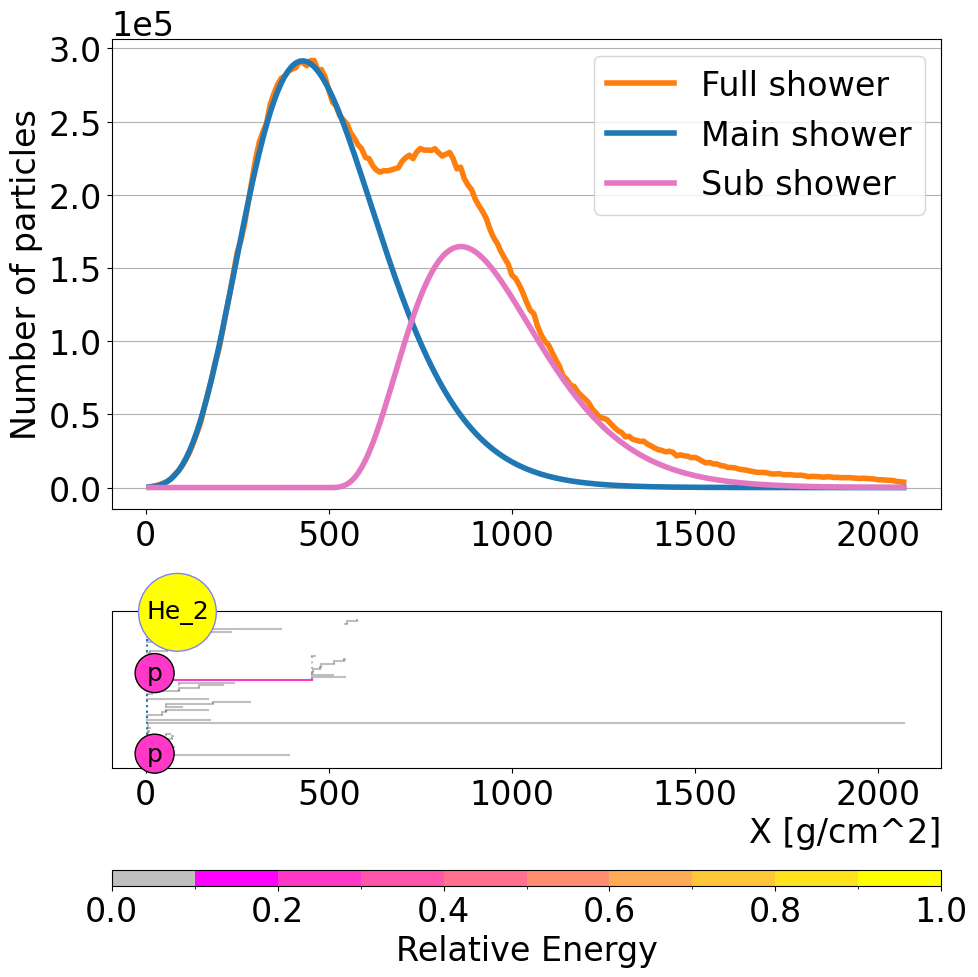}
	\includegraphics[width=0.49\columnwidth]{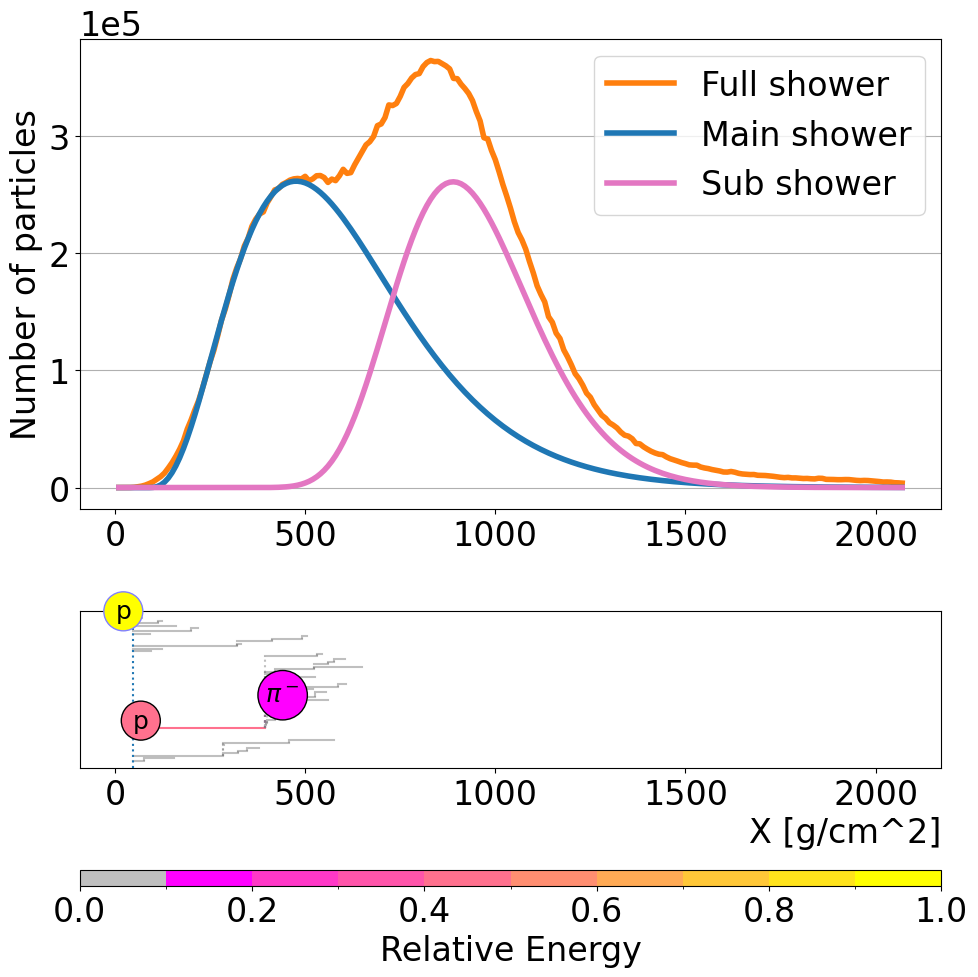}
    \caption{Two examples of double bump showers. The top panels show the number of particles as a function of atmospheric depth. The evolution of the full shower is fitted by a superposition of two Gaisser-Hillas profiles. The bottom panels show the underlying structure of the shower. Lines indicate the trajectories of the most energetic shower particles and labels indicate the particle type. Colors of the tracks and labels indicate the fraction of the primary cosmic-ray energy carried by the particle. The left (right) shower is produced by a 1 PeV Helium (proton) shower.}
    \label{fig:db_exaples}
\end{figure}

Figure \ref{fig:db_exaples} shows two examples of double bump showers simulated with CORSIKA v7.74~\citep{Corsika:1998}. The top plots show the longitudinal development of the shower. The orange lines show the number of shower particles versus the atmospheric depth. These anomalous profiles can be fitted with a superposition of two Gaisser-Hillas functions (the blue and magenta lines). The natural interpretation is that one of these components was initiated by a late interacting energetic particle.

To confirm this assumption, we can trace the particles in the CORSIKA simulation. The bottom panels in Fig.~\ref{fig:db_exaples} are \emph{skeleton plots} that reveal the structure of the shower. Each particle that carries more than 1\% of the primary energy is plotted as a horizontal line connecting its point of creation to the point where it has its next interaction. Grey lines represent particles with energies below 10\% of the primary energy, while the most energetic particles are brightly colored and given a label with a particle identifier. 

The left example is a shower created by a Helium primary. In the first interaction, the nucleus breaks apart in protons and neutrons. One of the protons travels more than 500 g/cm$^2$ before interacting and depositing its energy in the development of a secondary shower. In the right example, the primary particle is a proton. Approximately 50\% of the primary energy is concentrated in a single proton that travels more than 400 g/cm$^2$. 

To create a secondary bump, the long-flying particle must be energetic enough and cover a large enough distance. In our definition, double bump showers have longitudinal profiles for which a double Gaisser-Hillas function fits significantly better than a single function. To compare the fit quality of functions with a different number of free parameters we use the Akaike information criterion~\citep{Akaike:1974vps} with a likelihood ratio threshold of 0.01.

\begin{figure}
    \centering
	\includegraphics[width=0.32\columnwidth]{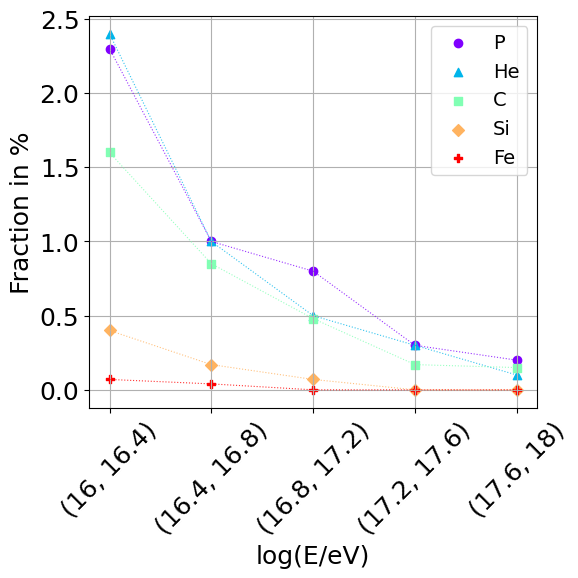}
    \includegraphics[width=0.32\columnwidth]{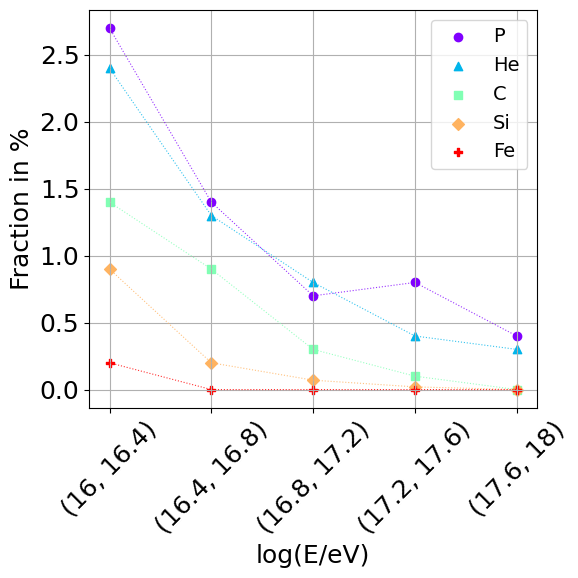}
    \includegraphics[width=0.32\columnwidth]{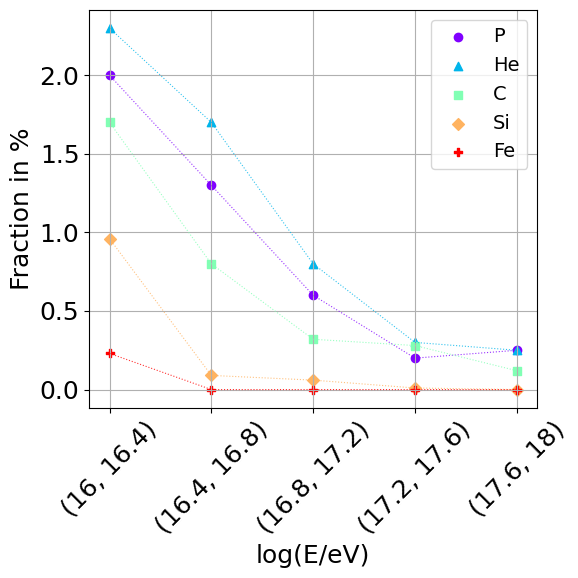}
    \caption{Fraction of showers that feature a double-bump structure as a function of energy for different primary particles and different hadronic interaction models: EPOS-LHC (left), Sibyll-2.3d (middle), and QGSJETII-04 (right). Each energy bin is based on 4000 simulated showers, so a fraction of 1\% corresponds to 40 double bump showers. Statistical fluctuations corresponding to these low quantities are visible in the plot.}
    \label{fig:fraction_of_db}
\end{figure}

Based on this definition, Fig.~\ref{fig:fraction_of_db} shows the fraction of simulated showers that has a double-bump profile for different primary mass and energy and using different hadronic interaction models. Although there are differences between the interaction models the overall picture is consistent. Double bumps occur most often for proton and Helium primaries. For protons the energy of the most energetic secondary hadron depends on the elasticity of the first interaction. For Helium double-bump showers, it is often one of the four nucleons that initiates the secondary shower component. In practice, these two scenarios lead to very similar probabilities for double-bump showers (see Fig.~\ref{fig:fraction_of_db}).

The fraction of double-bump showers decreases for heavier nuclei. The intuitive reason for this is that the primary energy of a heavy nucleus is divided over more nucleons. A secondary sub-shower created by one of these nucleons will contain less energy and is less likely to significantly alter the longitudinal evolution.

In addition, it can be seen that the fraction drops from 2-2.5\% to 0.25\% over the SKA-Low energy range (for light nuclei). The reason for this is that the hadronic cross section increases with energy and it becomes more unlikely for highly energetic particles to travel a large distance.

\subsection{Stretched showers}
When the secondary shower component is not far enough displaced from the main shower to produce a double bump, it may still have an impact on the longitudinal profile: it becomes elongated or stretched. In Eq.~\ref{eq:RL} this translates into a higher value for $L$. The probability distribution of the distance a particle travels before interacting is an exponential decay function. For each double bump shower, there must therefore exist many more stretched showers. 

\begin{figure}
    \centering
    \includegraphics[width=0.49\columnwidth]{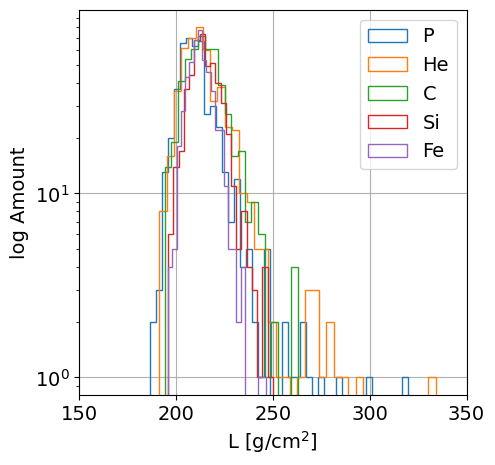}
    \includegraphics[width=0.47\columnwidth]{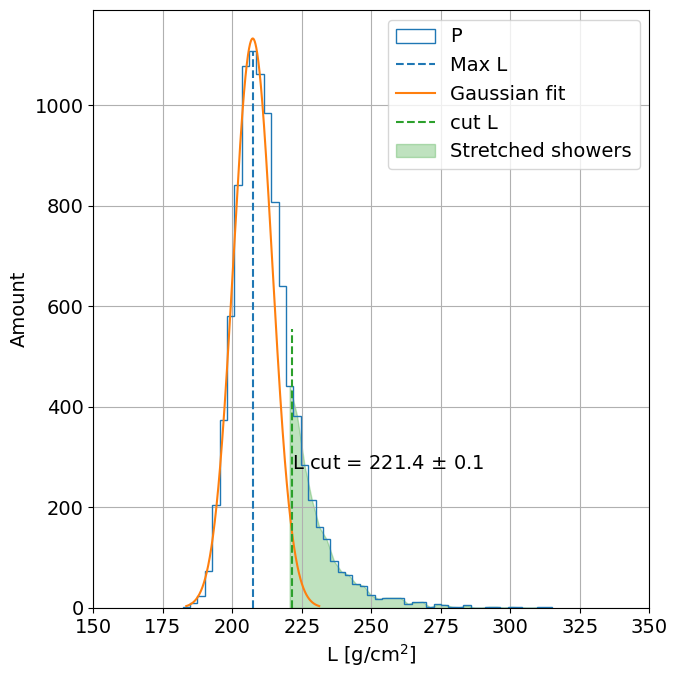}
    \caption{Left: Distribution of $L$ for sets of simulated 10-100 PeV showers of different primary mass. Double-bump showers were removed from the sample. Right: $L$-distribution of proton showers fitted with a Gaussian. Stretched showers are defined as having an $L$-value that is more than two sigma above the average value.} 
    \label{fig:L_histograms}
\end{figure}

The left panel of Fig.~\ref{fig:L_histograms} shows the distribution of the $L$ obtained when fitting `ordinary' showers (i.e. double bumps are removed from the sample). These distributions have the shape of a normal distribution with a tail containing high-$L$, or \emph{stretched} showers. This tail is most pronounced for proton and Helium, and almost absent for Iron. This follows the same trend as double-bumps showers, which is expected as they have a common origin. To count the number of stretched showers, we use the definition visualized in the right panel of Fig.~\ref{fig:L_histograms}.

\begin{figure}
    \centering
	\includegraphics[width=0.32\columnwidth]{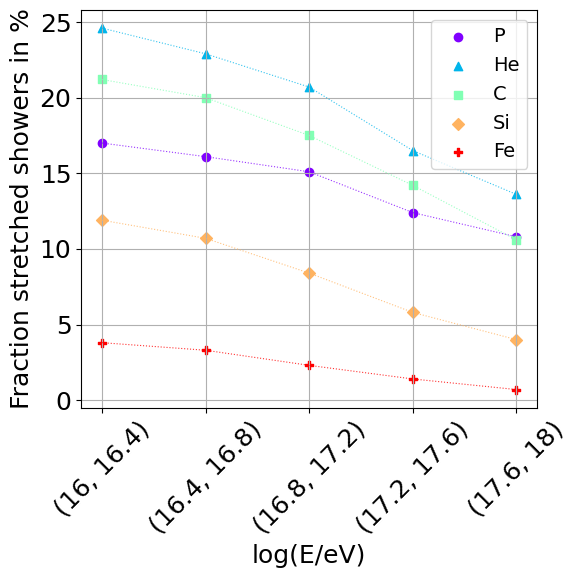}
    \includegraphics[width=0.32\columnwidth]{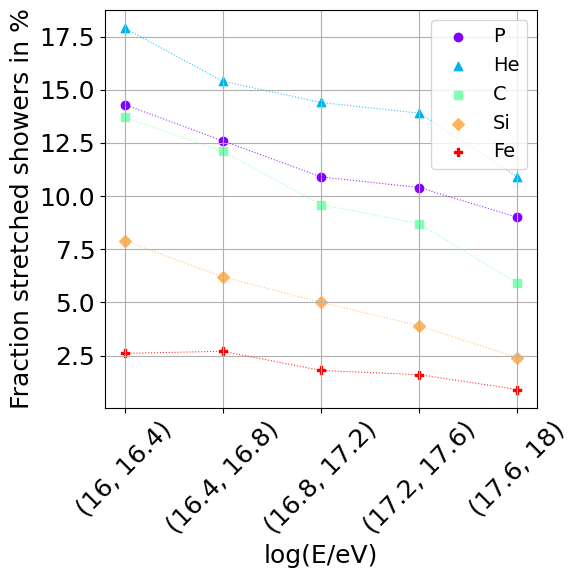}
    \includegraphics[width=0.32\columnwidth]{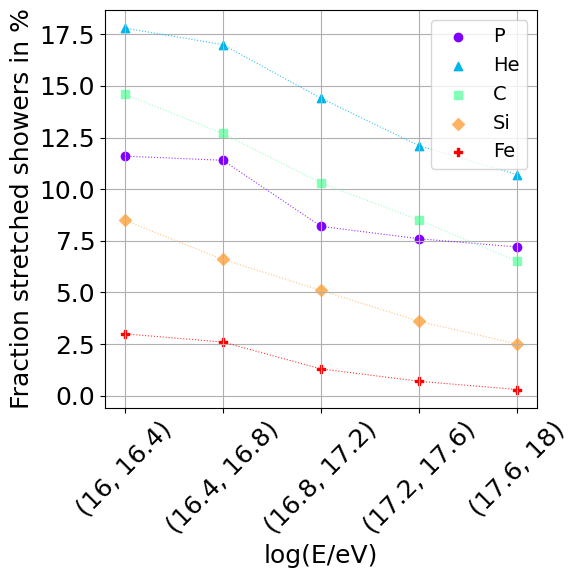}
    \caption{Fraction of stretched showers as a function of energy for different primary particles and different hadronic interaction models.}
    \label{fig:fraction_of_stretched}
\end{figure}

The number of stretched showers for different primary mass and energy and for different elements is plotted in Fig.~\ref{fig:fraction_of_stretched}. The dependence on mass and energy is similar to that of the double bump distributions (Fig.~\ref{fig:fraction_of_db}), but the fractions are larger by an order of magnitude.  

Double-bump and stretched showers are products of the same underlying principle: the deposit of a sizable fraction of energy late in the shower development. Whereas double-bump showers have the most dramatic features, they are less common than stretched showers. The relative frequency at which they occur depends critically on the primary and the hadronic cross sections. Therefore, the observation of anomalous showers will provide powerful new constraints on both the cosmic-ray mass composition and hadronic interaction models.    

\section{Simulations for SKA-Low}
The radio emission from air showers can be simulated with the CoREAS component of CORSIKA~\citep{CoREAS:2013}. The contribution to the emission of each electron and positron in the shower is computed for different observer positions on the ground. Such simulations are computationally very expensive and can take one or several days on a typical cluster node, depending on the shower energy and the number of antenna locations. Fortunately, it is not necessary to include all SKA antennas in the simulation setup. Instead, a fixed star-shaped pattern of observer locations is simulated (following the same prescription as \citep{Corstanje:2025}) from which the radiation at arbitrary position can be retrieved using a dedicated interpolator~\citep{Corstanje:2023vqp}. 

Since air showers propagate through the atmosphere at a highly relativistic velocity, the radiation pattern they emit is strongly beamed forward into a circle. The SKA-Low is mostly sensitive to air showers with a zenith angle below 60 degrees, for which the radiation patterns have a typical radius of 200-300~m. The emission mechanism itself is a combination of geomagnetic and charge excess radiation. The contributions combine into a complex shape that is not rotationally symmetric. The waveforms have a short time width of tens of nanoseconds. The integrated energy this pulse contains is called the \emph{fluence} (see \cite{Huege_2016} and references therein).

\begin{figure}
    \centering
	\includegraphics[width=1.0\columnwidth]{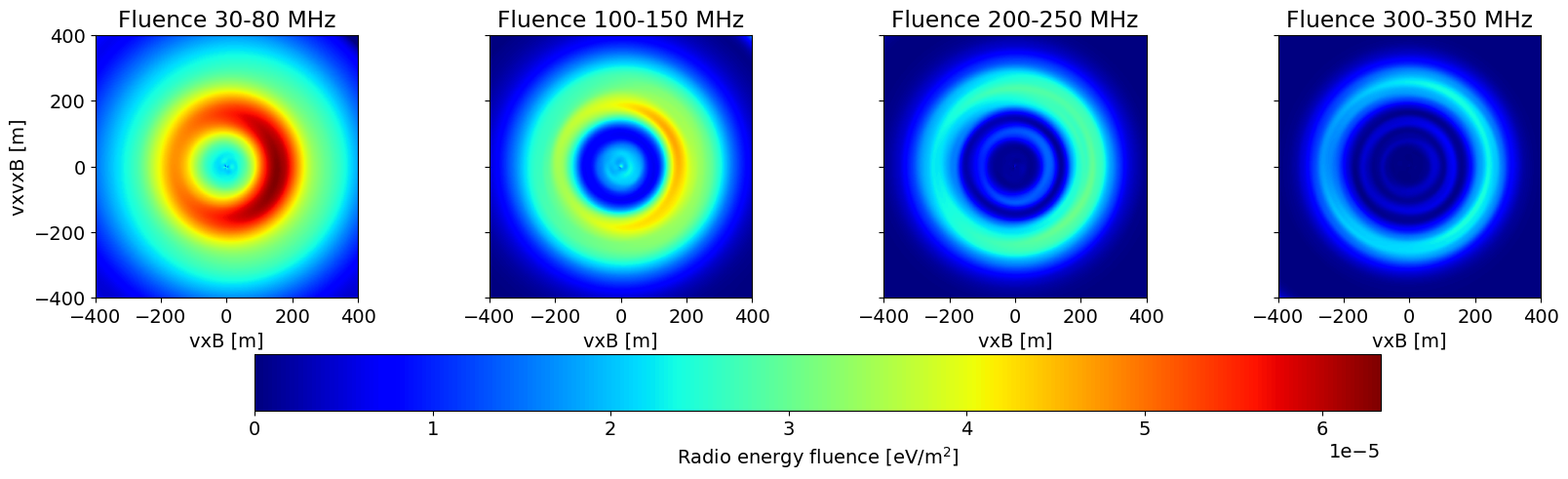}
    \caption{The fluence footprints of a 10 PeV Helium shower with a double-bump structure. The features of the pattern depend strongly on the selected frequency band.}
    \label{fig:Fluence_map_freqs}
\end{figure}

A helium double-bump shower with a primary energy of $5.6 \times 10^{17}$~eV was selected for full simulation. Fig~\ref{fig:Fluence_map_freqs} shows the fluence radiation pattern of this shower on the ground filtered to different frequency ranges. Above 100~MHz a secondary feature becomes visible in the center, while at even higher frequencies a multi-ring pattern emerges which is not the case for regular showers. This can be understood as an interference pattern between the two sub-showers. The structure of this pattern can be used to reconstruct the locations of the primary and secondary shower maximum, as we shall see below. 

\begin{figure}
    \centering
    \includegraphics[width=0.47\columnwidth]{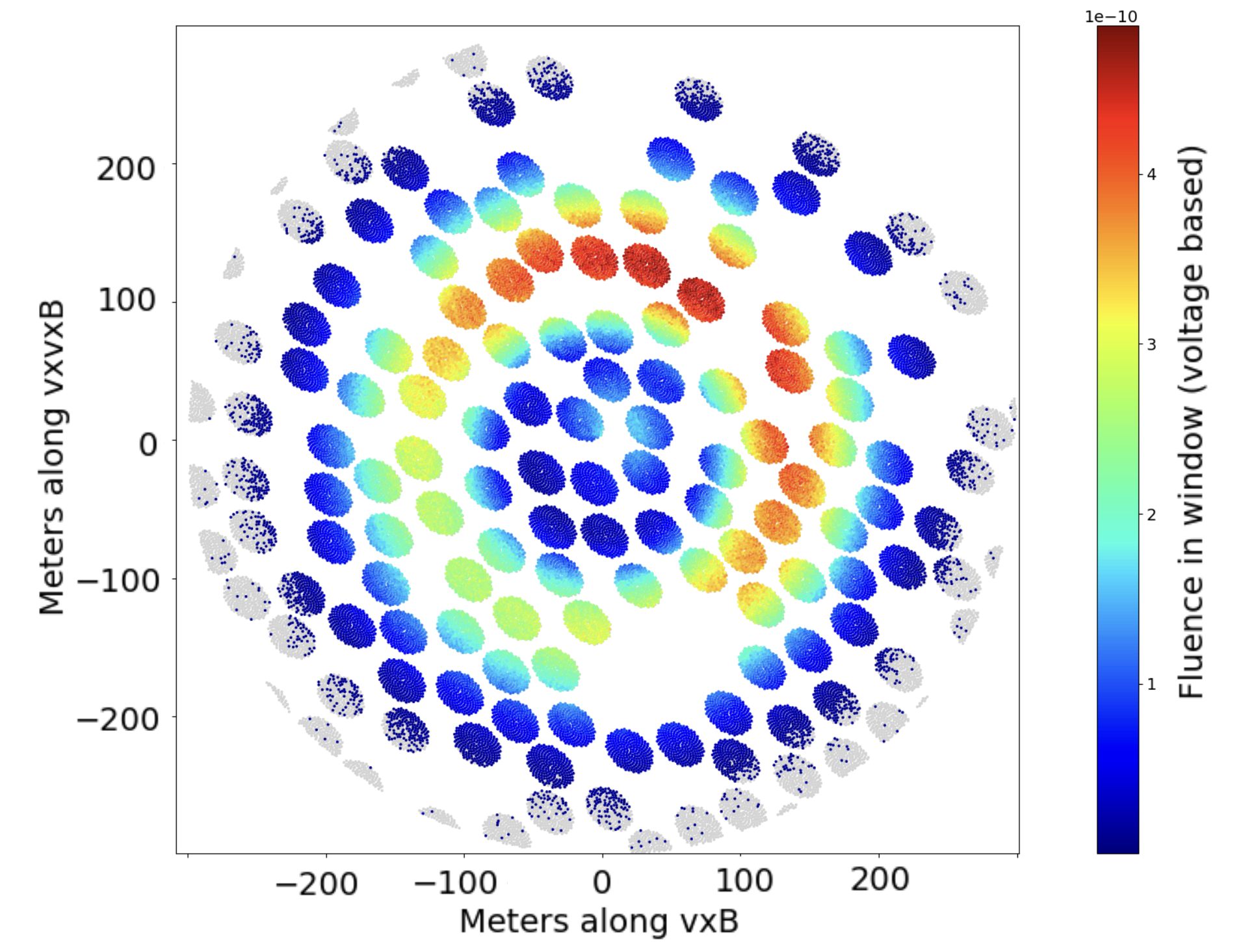}
    \includegraphics[width=0.52\columnwidth]{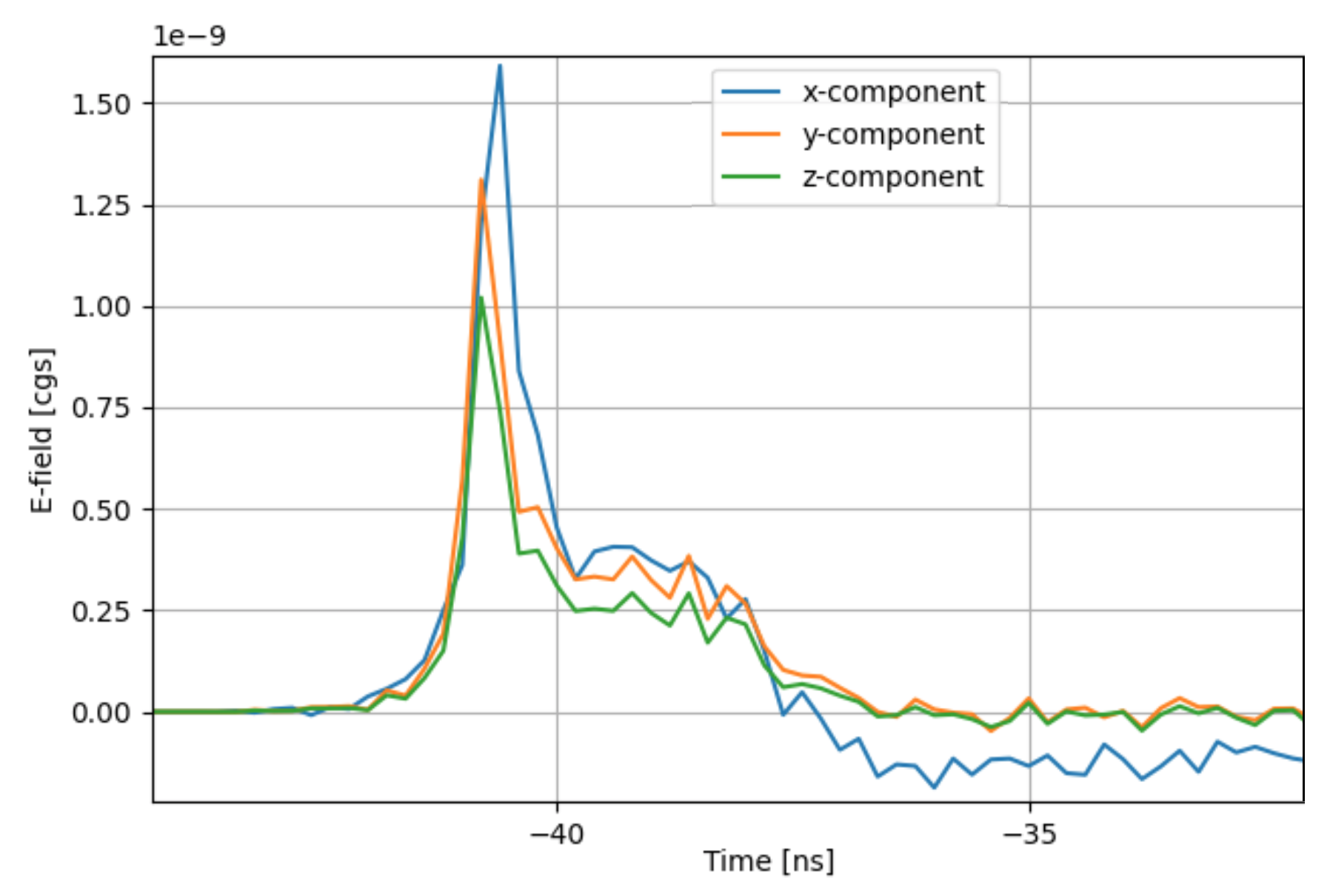}
    \caption{Left: simulations of a $5.6 \times 10^{17}$~eV helium shower. The color indicates the measured fluence per antenna including Galactic background noise and instrumental noise. Right: example waveforms for an antenna situated near the center of the shower footprint.}
    \label{fig:full_sim_SKA}
\end{figure}

To test whether these features can indeed be observed with the SKA we simulated the measured waveforms by applying the SKALA antenna model and adding realistic sky noise from the Galactic background and instrumental noise. The result for an example shower is shown in the left panel of Fig.~\ref{fig:full_sim_SKA}. It shows a helium shower with a primary energy of $5.6 \times 10^{17}$~eV, which is observed by thousands of antennas simultaneously. The main ring structure is similar to the radiation pattern of regular air showers, but the additional radiation near the center of the plot is unique for a double-bump shower. This excess radiation is observed with hundreds of showers and clearly recognizable by eye. The right panel shows an example of a radio pulse observed in this inner region and can be understood as the superposition of two components. The sharp peak is produced by the second bump: observers near the center view this component close to the Cherenkov angle, which results in a narrow peak with high-frequency components. Radiation from the main shower is similarly sharply peaked inside the large bright ring (at $\sim 150$~m radius) but has created a broad lower frequency pulse in the center. The resultant pulse shape is unique to double-bump showers.


\section{Reconstruction}

For regular air showers, there are several parameters that can be reconstructed from radio observations. The general geometry of the air shower is described by the core position where the shower axis hits the ground and the arrival direction. The energy of the primary cosmic ray can be inferred from the total radiation fluence integrated over the shower footprint. The cosmic-ray mass cannot be directly inferred, but it statistically correlates with \Xmax, the atmospheric depth of the shower maximum.

The exceptional resolution on \Xmax that SKA-Low can achieve by applying the reconstruction method used at LOFAR is demonstrated in \cite{Corstanje:2025}. In this approach, 20-30 different showers are simulated with CORSIKA/CoREAS. Each has the arrival direction and energy that matches the observation but a different value for \Xmax. For each simulation, the interpolated radiation pattern is fitted to the observations. The reconstructed value for \Xmax is then defined by the minimum of a parabola fit through the reduced $\chi^2$-values of the best-fitting simulations~\citep{Buitink:2014eqa}. 

The expected resolution of $5-8$~g/cm$^2$ outperforms all existing cosmic-ray observatories in the energy range. However, the technique is still only scratching the surface of what will be possible. For example, the observed signal was integrated over the full 50-350~MHz bandwidth, not using the additional information encoded in the shape of the frequency spectrum. Moreover, the exact shape of the radiation pattern is not fully determined by \Xmax alone, but depends on the complete shape of the longitudinal development of the air shower. For example, the $R$ and $L$ parameters in Eqn.~\ref{eq:RL} leave a unique fingerprint on the emission that can in principle be reconstructed with very precise observations. First hints of these effects were found with LOFAR, but so far no observatory was able to put meaningful constraints on the shape of individual showers~\citep{Buitink:ARENA24}. SKA-Low will be the first telescope to achieve this, due to its extremely high antenna density and broad bandwidth.
A description of the progress on simultaneously reconstructing \Xmax, $R$, and $L$ can be found in \cite{Corstanje2026.SKA}.

Several approaches are possible to reconstruct the longitudinal shower structure. Following the procedure described above, the radiation of showers with different longitudinal profiles can be fitted directly to the data. Although the required computational effort seems excessive when using traditional Monte Carlo simulations (CORSIKA/CoREAS), faster codes like SMIET~\citep{Desmet:2025ufy} and MGMR3D~\citep{Scholten:2018} make this a viable approach. Such codes can be used for reconstruction strategies similar to the LOFAR \Xmax reconstruction, or more sophisticated strategies using the full waveform information, e.g. in the framework of Information Field Theory~\citep{Watanabe2026.SKA}.  

Alternatively, interferometric techniques could be used to directly image the charge-current distribution of the shower. However, there is no mathematically robust algorithm for interferometry in the near-field for a relativistically moving system. In the RIT technique~\citep{Schoorlemmer:2020low}, signals from all antennas are coherently added (beamforming) after applying time delays corresponding to emission points along the shower axis. The point of maximum emission found with this technique strongly correlates with the shower maximum. Alternatively, the full longitudinal evolution of the transverse current can be reconstructed by using a more formal approach~\citep{Scholten:2024upn}.


Double-bump showers pose an additional challenge to reconstruction algorithms. As shown in the previous Section, both the radiation pattern and pulse waveforms are very different from regular showers. However, some basic geometrical arguments can be used to get a good estimate of the double bump properties.

The radiation observed at a specific antenna position can be described as the superposition of the contributions from the primary and secondary shower. We can model these two signals as coming from the two peaks in the longitudinal profile $X_{\rm max1}$ and $X_{\rm max2}$. The time difference $\Delta T$ between these two signals can be calculated, taking into account the slight but important difference between the velocities of the shower particles (close to speed of light in vacuum) and the radio waves (speed of light in air). At the frequency $\omega = \pi / \Delta T$ the two contributions interfere destructively. 

\begin{figure}
    \centering
	\includegraphics[width=0.49\columnwidth]{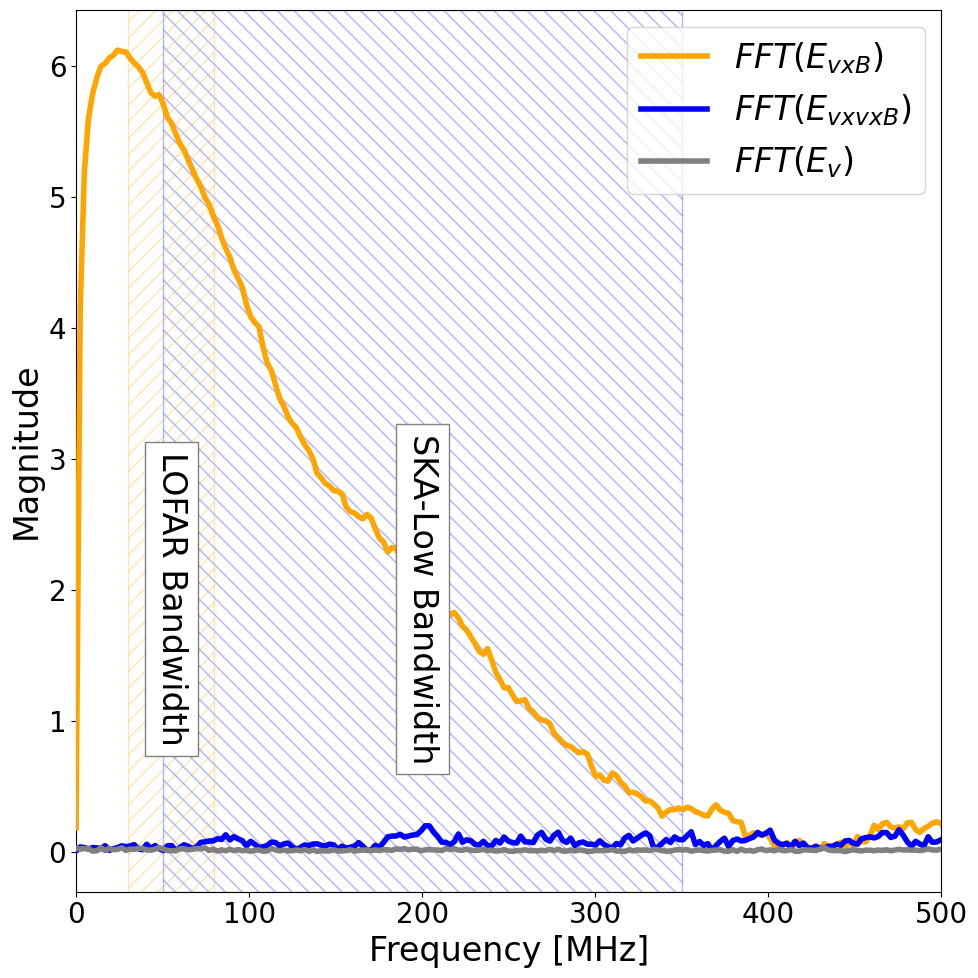}
	\includegraphics[width=0.49\columnwidth]{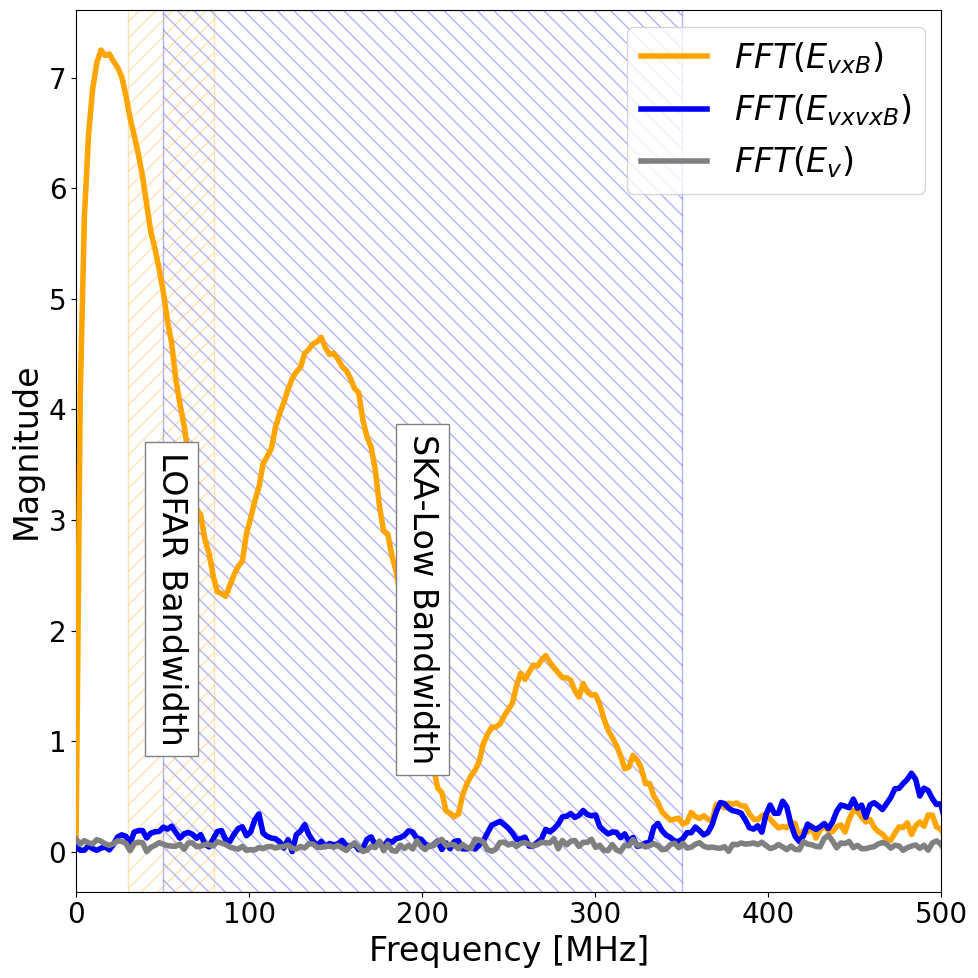}
    \caption{Frequency spectra of a radio pulse from a regular shower (left) and a double-bump shower (right). The interference maxima of the double peak spectra usually fall outside of the LOFAR band but inside the SKA-Low band.}
    \label{fig:freq_interference}
\end{figure}

This effect is illustrated in Fig.~\ref{fig:freq_interference}. The left panel shows the frequency spectrum of a regular air shower, while the interference pattern from a double bump shower emerges clearly in the spectrum on the right. The first interference minimum is almost exactly at the frequency predicted by the point-source model. Secondary maxima and minima deviate more from the predictions, presumably because the assumption that all emission is coming from the two shower maxima breaks down. 

The frequency of interference minimum depends on the geometry of the shower and the antenna position, but usually falls in the $75-250$~MHz range. As indicated in the Figure, this falls outside the $30-80$~MHz range of LOFAR and the Pierre Auger Observatory. SKA-Low, on the other hand, covers the complete range of the interference pattern.

\begin{figure}[h]
    \centering
	\includegraphics[width=0.9\columnwidth]{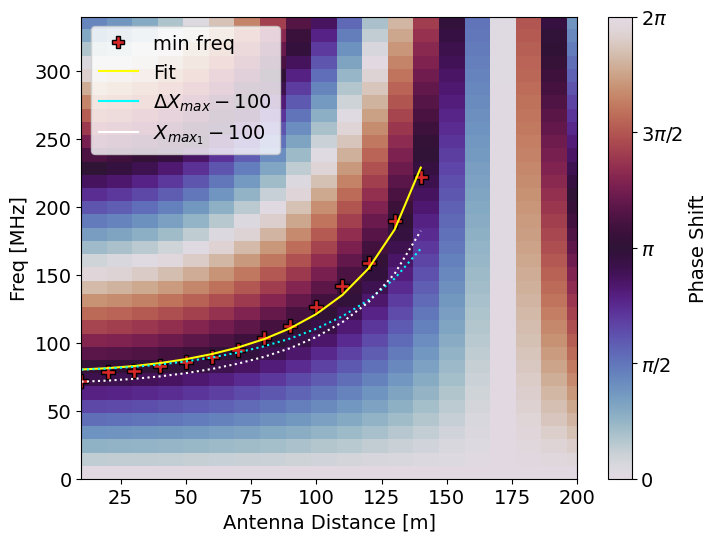}
    \caption{Double-bump reconstruction plot. The background colors indicate the relative phase of the radio signals from the two sub-showers based on the point-source model for different observer positions and frequencies. The red crosses are the first interference minima found in the spectra of fully simulated waveforms. The yellow line is a fit of the data to the point-source model. Two additional model lines are shown for reference.}
    \label{fig:db_reconstruction}
\end{figure}

Fig.~\ref{fig:db_reconstruction} demonstrates a shower reconstruction method based on the point-source model. The background colors indicate the phase shift between the two emission contributions as a function of antenna distance and frequency. At a distance of $r\sim 170$~m, $\Delta T=0$ and the interference is constructive at all wavelengths. The first interference minimum appears as the dark fringe in the center of the plot. The red crosses indicate the frequencies at which a minimum was found in the spectrum of individual antennas. These points can be fitted with the point-source model (yellow line) to find $X_{\rm max1}$ and $X_{\rm max2}$. For reference, two alternative lines are given that correspond to models in which $X_{\rm max1}$ or $\Delta$\Xmax is shifted by 100~g/cm$^2$. 

The point-source model is useful for understanding the emission from double-bump showers and to perform a preliminary reconstruction of $X_{\rm max1}$ and $X_{\rm max2}$. These values can then be used as an ansatz in a more sophisticated analysis incorporating full signal simulations. The performance of such reconstructions is still being studied. The final reconstruction will not only yield $X_{\rm max1}$ and $X_{\rm max2}$, but also the energies $E_1$ and $E_2$ of the two sub-showers as well as the total electromagnetic energy of the shower.     

\section{Analysis}

This section is based on a set of CONEX simulations at three fixed energies (10 PeV, 100 PeV, 1 EeV) and for five elements (p, He, C, Si, Fe) using three hadronic interaction models (EPOS-LHC, Sibyll 2.3d, and QGSJETII-04). Ten thousand showers were generated for each combination of energy, mass, and interaction model. 

\subsection{Particle identification from energy ratio}

\begin{figure}[h]
    \centering
	\includegraphics[width=0.49\columnwidth]{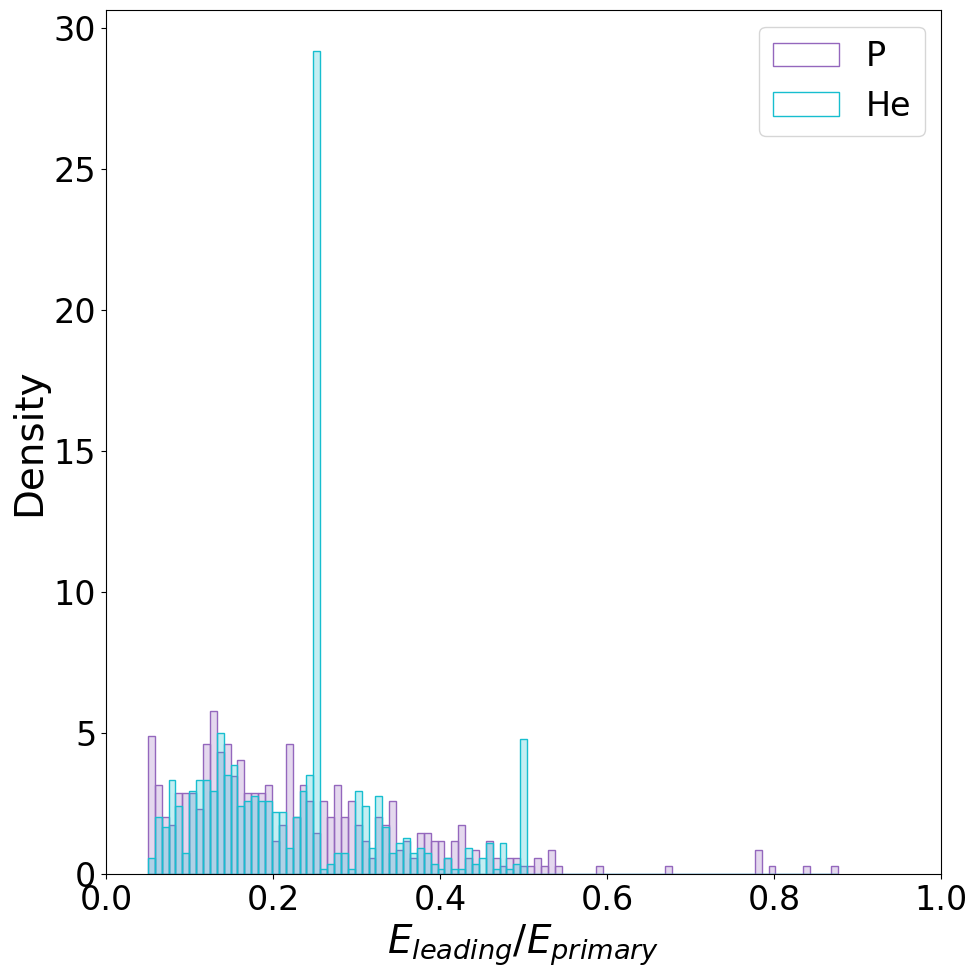}
    \includegraphics[width=0.49\columnwidth]{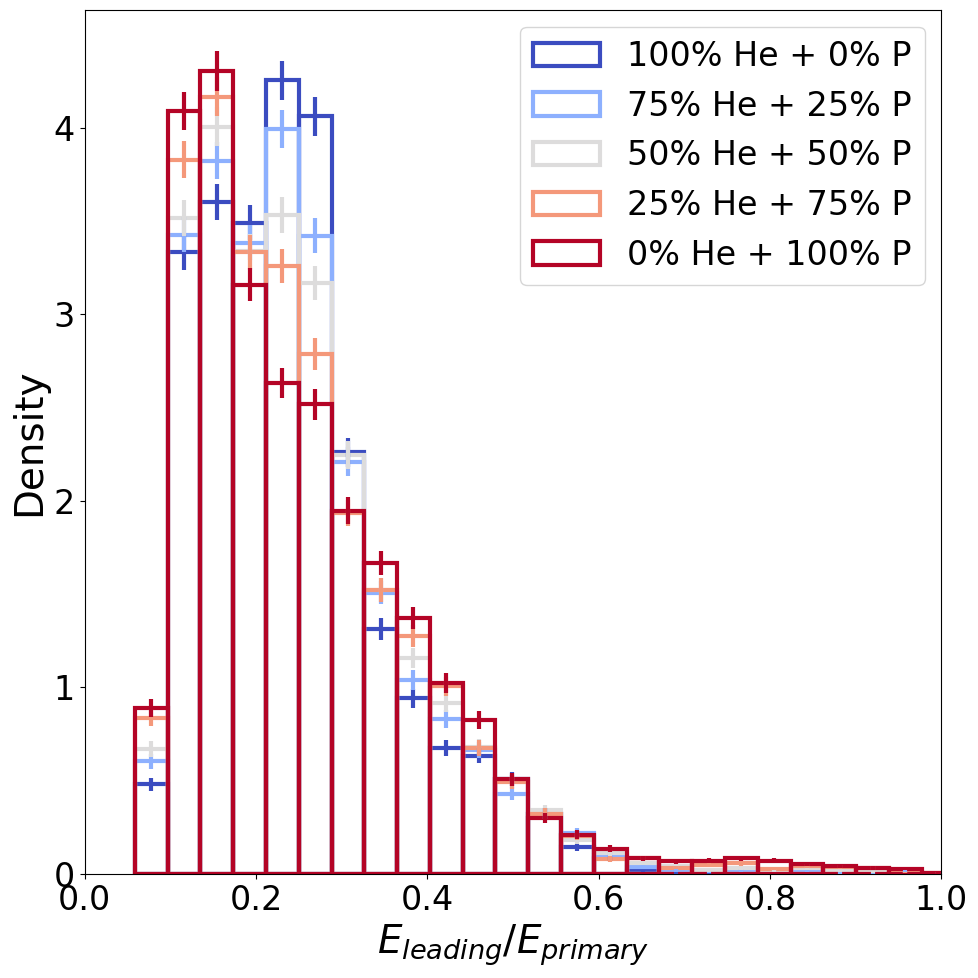}
    \caption{Left: distribution of energy ratio between the secondary bump and the total shower for helium and proton. Double bump showers with a ratio below 0.1 are excluded from the analysis. Right: Energy-ratio distribution for sets of ten thousand cosmic rays drawn from fluxes with different p-He ratios. The energy contained in the secondary bump ($E_{\mathrm{leading}}$) is divided by the energy of the primary cosmic ray. Gaussian uncertainties in the energy resolution have been added as with $\sigma=10$\%.}
    \label{fig:energy_ratios}
\end{figure}

The ratio between the energy contained in the two parts of a double bump shower can be used to constrain the mass of the primary particle. In many cases, the secondary shower is initiated by a hadron of energy $E_{tot}/A$, with $A$ being the mass number. This occurs when one of the nucleons in the primary cosmic ray retains its momentum when the nucleus breaks up in the first interaction or when it transfers almost all of its momentum to a secondary hadron. When this process occurs in a Helium shower, the secondary shower will have close to 25\% of the total energy. The left panel of Fig.~\ref{fig:energy_ratios} shows the distribution of the ratio between the energy contained in the second bump and the total energy. For helium, it is continuous with a sharp peak at 25\%. A second smaller peak is seen at 50\%, corresponding to showers in which (a fragment consisting of) two nucleons initiate the second bump. 

The reconstruction resolution of energy ratios is still under investigation. Since the ratio has a strong impact on the shape of the interference radiation pattern on the ground, a resolution of 10\% is a reasonable expectation. The right panel of Fig.~\ref{fig:energy_ratios} shows that the Helium peak is still visible when Gaussian uncertainties of this level are included for a total number of detected double bump showers of 10,000. It can thus be used as a unique identifier of a Helium component in the cosmic-ray flux. The position of the peak is independent from the choice of hadronic interaction model because it follows directly from the number of nucleons in a Helium atom.  

The detection rate will depend on the maximum triggers per hour that will be possible at SKA-Low without interfering with other ongoing observations~\citep{Corstanje:2025}. If we assume a low trigger rate of $\sim$10 per hour, thousands of double bumps can be detected per year, provided that there is a strong light component in the cosmic-ray flux.    

Heavier nuclei produce similar peaks in the energy ratio, but since the fractional energy is much smaller (below 10\% for elements in the CNO group), this will be much harder to observe.

For proton showers, another unique signal can occur. Since the hadron that initiates the secondary component in a proton double-bump shower can have any fraction of the primary particle, it can occur that it carries more than half of the total energy. In this case, the first bump in the shower profile is smaller than the second bump. The energy ratio distribution of proton double-bumps in Fig.~\ref{fig:energy_ratios} is smooth and extends to values above 50\%. 

\subsection{Hadronic cross sections from peak separation}

\begin{figure}
    \centering
	\includegraphics[width=0.9\columnwidth]{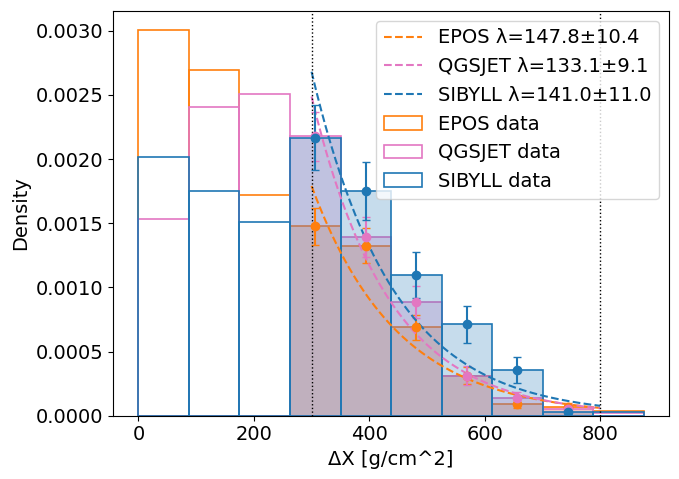}
    \caption{Distribution of the peak separation $\Delta X$ for a set of 10 PeV double-bump showers for different hadronic interaction models. For each model, an exponential fit is performed including only values $\Delta X > 300$~g/cm$^2$}
    \label{fig:peak_separation}
\end{figure}

The distribution of the peak separation $\Delta X$ follows from the hadronic cross section in the simulation and can thus be used to test the predictions of different hadronic interaction models or even directly measure the hadronic cross section. Figure \ref{fig:peak_separation} shows the $\Delta X$ distribution for double bumps found in simulations of 10 PeV showers. Only showers with a clearly identifiable double-peak structure are included, which causes the suppression at low $\Delta X$ values. Above $\Delta X \sim 300$~g/cm$^2$ the recognition of the double-bump structure becomes close to 100\% and the distribution can be fitted by an exponential function. To first order, the exponential coefficient $\lambda$ can be used to directly infer the interaction cross section of a particle. However, there are complicating factors that make the interpretation of $\lambda$ more subtle.

First of all, the simple picture in which a single hadron travels the complete distance $\Delta X$ without interacting only accounts for some of the double bumps. Other scenarios include a combination of two or more long tracks. The hadron interacts in-between these segments but transfers most of its energy to the leading particle which again travels far. Alternatively, the total energy in the second bump can be provided by two or more particles that travel in parallel. We refer to these scenarios as \emph{ladder events} and \emph{branching events} and they form a large contribution to the total set of double bump showers. 

Secondly, the distance between the two maxima in the longitudinal profile will not match exactly to the distance that the hadron has traversed because it also depends on the shape of the longitudinal development of the two sub-showers. Finally, it is not possible to distinguish protons, neutrons, pions, and possibly larger nuclear fragments. 

Consequently, the measured $\Delta X$ will depend not only on the cross section but also on other interaction properties such as multiplicity and elasticity. The $\lambda$ coefficient obtained from observations should therefore be regarded as an \emph{effective interaction length} and compared directly with predictions from different hadronic interaction models. The fits in Fig.~\ref{fig:peak_separation} demonstrate the difference between the models. Alternatively, the properties of hadronic interactions inside a single model can be modified with a scaling parameter to study how each property individually changes the observed double bump distribution, similarly to the approach of \cite{Ulrich:2010rg}.      

\subsection{Combined analysis of stretched showers and average Xmax}

Stretched showers can be analysed in a manner similar to double-bump showers. The fraction of stretched showers can be determined and compared to predictions of different interaction models. Thus, a combined analysis provides a check on the reconstruction performance and can be used to estimate systematic uncertainties.

\begin{figure}
    \centering
	\includegraphics[trim={1cm 0 1.2cm 0},clip,width=0.45\columnwidth]{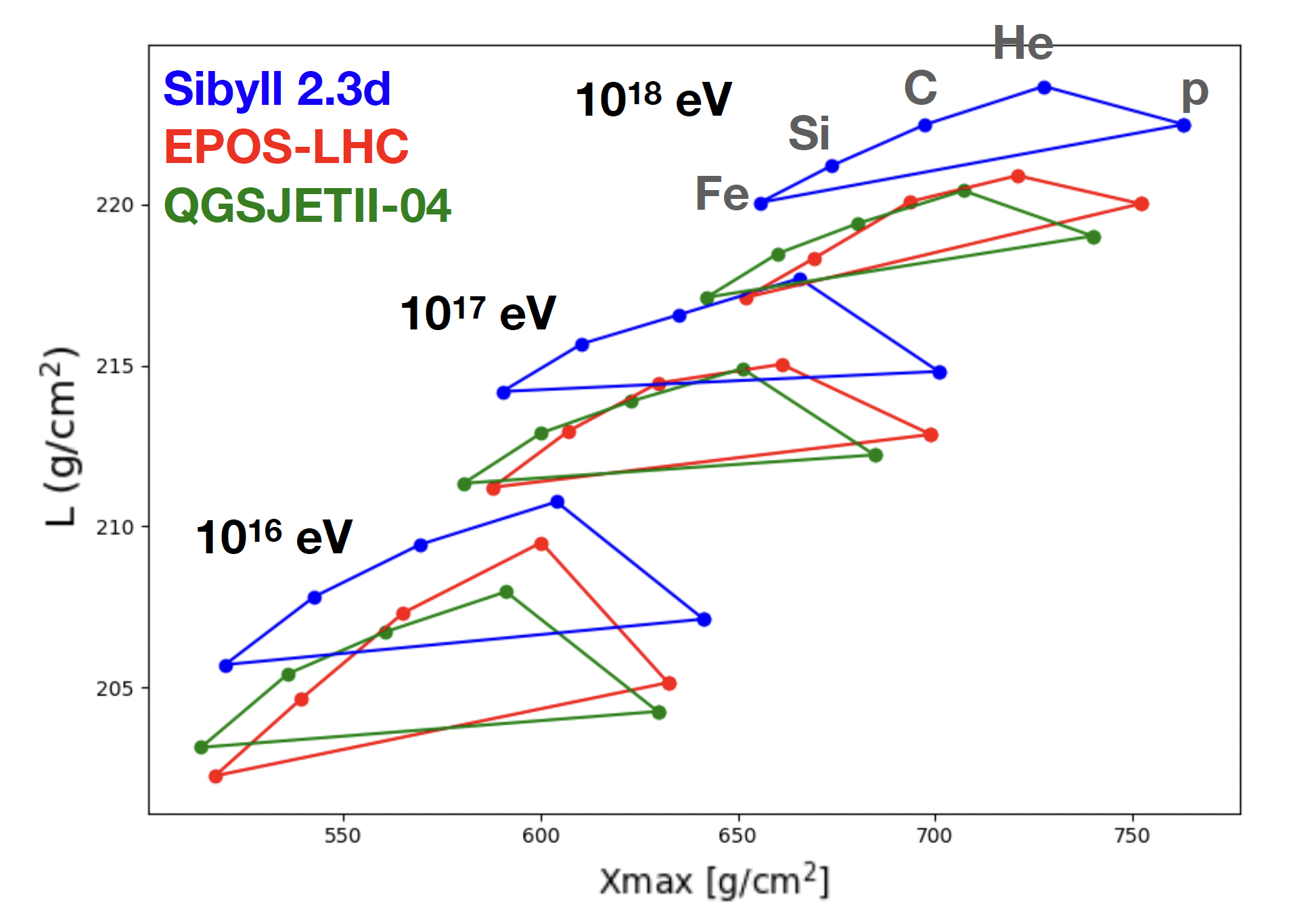}
    \includegraphics[trim={1cm 0 3.5cm 0},clip,width=0.54\columnwidth]{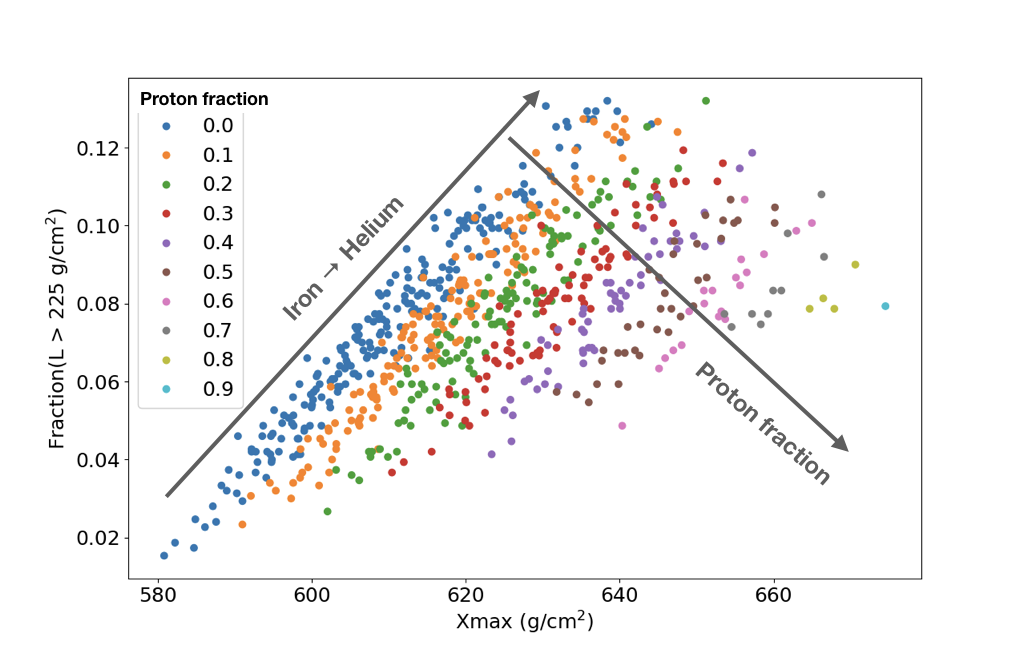}
    \caption{Left: average \Xmax and $L$ for showers of different energies, primary masses, and simulated with different hadronic interaction models. The dots represent samples of pure composition. Right: a strategy to separate protons from other particles. Each dot represents a unique mixed composition of 10$^{17}$ eV showers simulated with EPOS-LHC. Mixtures separate into diagonal bands according to their proton fraction.   }
    \label{fig:Xmax_L}
\end{figure}

At the higher end of the SKA-Low energy spectrum (above 10$^{17}$~eV) double-bump showers become rare, but stretched showers still occur frequently (see Fig.~\ref{fig:fraction_of_stretched}). For each observed shower, the longitudinal profile can be fitted to the data to simultaneously obtain \Xmax and $L$ (see also \cite{Corstanje2026.SKA} and \cite{Watanabe2026.SKA}). Figure \ref{fig:Xmax_L} demonstrates the power of obtaining this additional information.  

The left panel shows the average values of \Xmax and $L$ for different primary masses, energies, and hadronic interaction models. The points labeled with the names of the elements represent a pure composition of only that element. Since the real cosmic-ray flux is a mixture of elements, it will lie somewhere in the middle of the triangle, or wedge, that these points form (shown by connecting lines). 

In previous observations, only \Xmax is available. Depending on the choice of interaction model, the interpretation of the observed average \Xmax will change. For example, for a given average \Xmax at $10^{18}$~eV, Sibyll 2.3d will suggest a heavier composition than the other two models. By combining observations of \Xmax and $L$ this confusion is reduced. For a consistent model, the observations should fall within a model triangle at each energy. 
This analysis approach can be made more robust by considering the full shape of the distribution of observed \Xmax and $L$ instead of only the average values.

An interesting characteristic of $L$ is that the highest average values are found for helium. This can be understood from the distributions plotted in Fig.~\ref{fig:L_histograms}. Helium primaries have the highest probability of producing stretched showers. The longer tail in the distribution shifts the average value. In practice, counting the showers in the tail is a more robust parameter than the average of $L$. 

When combining \Xmax and $L$, protons have a unique characteristic. For all elements between iron and helium, there is a monotonous trend towards higher \Xmax and $L$. However, this trend is broken for protons, which have a higher average \Xmax than helium showers, but a lower $L$. This feature can be used to separate protons from the other elements. 

The right panel of Fig.~\ref{fig:Xmax_L} demonstrates this separation for EPOS-LHC showers of 10$^{17}$~eV. Each dot represents a thousand showers that are randomly drawn from the CONEX simulation sets and has a unique mix of proton, helium, carbon, silicon, and iron. From these thousand showers, the average \Xmax and the number of showers with $L>225$~g/cm$^2$ are calculated and plotted as a value with a color code that shows its proton fraction. 

Compositions with a fixed proton fraction form separate diagonal bands. This information can be used to directly measure the proton fraction. Note that other elements cannot be separated in a similar way. For example, a point halfway between helium and iron could be a mixture of those two, pure oxygen, or many other combinations of elements.    

Separating the protons from the all-particle cosmic-ray spectrum makes it possible to identify the transition from galactic to extragalactic flux. The highest-energy Galactic cosmic-rays are likely heavy elements that can be accelerated more efficiently in shocks. The lowest energy extragalactic cosmic rays, on the other hand, are expected to be dominantly protons.

\section{Conclusion}
SKA-Low will perform cosmic-ray observations in the Galactic-to-extragalactic transition region with an unprecedented precision. With current state-of-the-art techniques, it is only possible to reconstruct the atmospheric depth of the shower maximum \Xmax. SKA-Low will be the first observatory that can also measure the shape of the shower evolution for individual showers. In particular, it can search for showers with an anomalous evolution, such as double-bump and stretched showers. The rate at which such showers occur and their characteristics can be used to study and constrain hadronic interaction models. In addition, they provide new tools to determine the cosmic-ray mass composition. SKA-Low will be a cosmic-ray observatory with unique capabilities and can play an important role in understanding the astrophysics of the highest-energy Galactic sources and the onset of the extragalactic component.

\section*{Acknowledgments}
The authors build on countless efforts to enable air shower observations using radio emission and are indebted to the community. Concretely, we acknowledge the following support: 
SBo, AN, and KT acknowledge the Verbundforschung of the German Ministry for Research, Technology and Space (BMFTR). 
PL and KW are supported by the Deutsche Forschungsgemeinschaft (DFG, German Research Foundation) – Projektnummer 531213488.
BH, CS, and PT are supported by ERC Grant Agreement No. 101041097
ST acknowledges funding from the Khalifa University RIG-S-2023-070 grant.
KM acknowledges funding from the Netherlands Research School for Astronomy (NOVA) and Dutch Research Council (NWO) project OCENW.XS25.1.237. This research is supported by the Flemish Foundation for Scientific Research (FWO-AL991 and FWO-OZR4291)
The authors gratefully acknowledge the computing time provided on the high-performance computer HoreKa by the National High-Performance Computing Center at KIT (NHR@KIT). This center is jointly supported by the Federal Ministry of Education and Research and the Ministry of Science, Research and the Arts of Baden-Württemberg, as part of the National High-Performance Computing (NHR) joint funding program. HoreKa is partly funded by the German Research Foundation.

\bibliographystyle{abbrvnat-maxbibnames4}
\bibliography{chapter} 

\end{document}